\documentclass[sigplan,nonacm]{acmart}
\AtBeginDocument{%
  }

\setcopyright{none}

\acmConference[arXiv]{}{}{}
\usepackage{algorithmic}
\usepackage{textcomp}

\usepackage{subcaption}
\usepackage{makecell}
\usepackage{listings}
\usepackage{float}
\usepackage{stfloats}
\usepackage{tabularx}
\usepackage{multirow}
\usepackage{fancyvrb}

\usepackage{comment}
\usepackage{balance}
\usepackage{xspace}
\usepackage{cancel}
\usepackage{soul}
\usepackage{enumitem}
\usepackage{ulem}
\usepackage{pifont}

\lstdefinestyle{mystyle}{
    basicstyle=\ttfamily\footnotesize,
    breakatwhitespace=false,
    breaklines=true,
    keepspaces=true,
    showspaces=false,
    showstringspaces=false,
    showtabs=false,
    tabsize=2,
    numbers=left,
    numbersep=5pt,
    captionpos=t,
    frame=shadowbox,
    rulesepcolor=\color{gray!30}
}
\newcommand{\FM}{{\sc IMPL-NAME}\xspace}

\newcommand{\NAME}{\texttt{TECH-NAME}\xspace}

\newcommand{\zzc}[2]{{#1}} 

\begin{document}

\title{Partial Cross-Compilation and Mixed Execution for Accelerating Dynamic Binary Translation}

\author{Yuhao Gu}
\orcid{0009-0009-8761-1237}
\affiliation{%
  \institution{Sun Yat-sen University}
  \city{Guangzhou}
  \country{China}
}
\email{guyh9@mail2.sysu.edu.cn}

\author{Zhongchun Zheng}
\orcid{0009-0002-8176-8486}
\affiliation{%
  \institution{Sun Yat-sen University}
  \city{Guangzhou}
  \country{China}
}
\email{zhengzhch3@mail2.sysu.edu.cn}

\author{Nong Xiao}
\authornote{Corresponding author.}
\orcid{0000-0002-2166-977X}
\affiliation{%
  \institution{Sun Yat-sen University}
  \city{Guangzhou}
  \country{China}
}
\email{xiaon6@mail.sysu.edu.cn}

\author{Yutong Lu}
\orcid{0000-0001-5315-3375}
\affiliation{%
  \institution{Sun Yat-sen University}
  \city{Guangzhou}
  \country{China}
}
\email{luyutong@mail.sysu.edu.cn}

\author{Xianwei Zhang}
\authornotemark[1]
\orcid{0000-0003-3507-4299}
\affiliation{%
  \institution{Sun Yat-sen University}
  \city{Guangzhou}
  \country{China}
}
\email{zhangxw79@mail.sysu.edu.cn}

\begin{abstract}

With the growing diversity of instruction set architectures (ISAs), cross-ISA program execution has become common. Dynamic binary translation (DBT) is the main solution but suffers from poor performance. Cross-compilation avoids emulation costs but is constrained by an "all-or-nothing" model—programs are either fully cross-compiled or entirely emulated. Complete cross-compilation is often unfeasible due to ISA-specific code or missing dependencies, leaving programs with high emulation overhead.

We propose a hybrid execution system that combines compilation and emulation, featuring a selective function offloading mechanism. This mechanism establishes cross-environment calling channels, offloading eligible functions to the host for native execution to reduce DBT overhead. Key optimizations address offloading costs, enabling efficient hybrid operation. Built on LLVM and QEMU, the system works automatically for both applications and libraries. Evaluations show it achieves up to 13x speedups over existing DBT, with strong practical value.

\end{abstract}

\maketitle

\section{Introduction}

With the slowdown of Moore's Law\cite{chrisGoldenAgeCompiler2021, hennessyNewGoldenAge2019}, we have witnessed the rapid proliferation of diverse instruction set architectures (ISAs), such as x86, ARM and RISC-V.
As a result, the demand for cross-ISA execution has become increasingly prevalent. Emulators and dynamic binary translation (DBT) have emerged as the mainstream solutions, widely adopted to support legacy applications\ and port applications to new ISAs\cite{wenzlHackElaborateTechnique2019}, such as Rosetta 2 for MacOS\cite{Rosetta2Mac2021} and Prism for Windows on Arm\cite{mattwojoHowEmulationWorks2024}. However, DBT consistently faces severe performance challenges:
since a single guest\footnote{Throughout this paper, we use "guest" to refer to the side of the program being emulated, and "host" to denote the side of the native system that performs the emulation.} instruction often requires translation into multiple host instructions, and further runtime translation itself incurs significant overhead, emulated execution can be dozens of times slower than native execution \cite{zhangHermesFastCrossISA2015, liCrossDBTLLVMBasedUserLevel2022, wangGeneralPersistentCode2016}.

Meanwhile, most machine code executed at runtime is generated by compiling from high-level languages (e..g. C/C++)  rather than written directly in assembly, which implies a high degree of target-independency.
By cross-compiling\footnote{Conventionally, cross-compilation refers to compiling and generating guest-ISA programs on the host-ISA machines. Herein, we generalize its meaning to denote building the project source code originally written for a legacy ISA into binaries targeting a different new ISA.}
the source code to the host ISA, one can obtain a native executable that completely eliminates the interpretation overhead of DBT.
Yet, existing approaches confined to an "all-or-nothing" paradigm: either the entire program can be cross-compiled and executed natively, or cross-compilation fails, forcing the whole program to fall back to DBT emulation.
In practice, complete cross-compilation is often infeasible. For instance, the source code may include platform-specific macros or assembly codes, relying on middleware libraries unavailable on the host ISA, or simply lack accessible source code — such as legacy applications and commercial closed-source software. In such scenarios, it becomes unavoidable to endure the heavy overhead of full DBT emulation.

\begin{figure}[!h]
    \centering
    \includegraphics[width=0.85\linewidth]{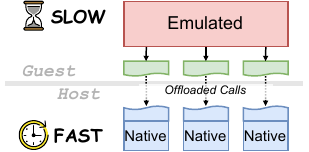}
    \caption{\NAME mechanism enables partial offloading of guest code to the host side for native execution, thereby bypassing the emulation overhead of DBT.}
    \label{fig:intro}
\end{figure}

This paper presents \NAME, a function offloading mechanism that enables cross-compilation and hybrid code execution for different guest and host ISAs to accelerate DBT emulation, as shown in Figure \ref{fig:intro}.
Such offloading require supports from both compile-time and runtime.
At compile-time, \NAME extracts eligible guest functions and generates semantically equivalent host functions. At runtime, guest-side calls to these functions are forwarded to the host side for native execution, thereby bypassing the substantial emulation overhead of DBT.
\NAME successfully addresses two key challenges in offloading: first, the discrepancy between guest and host ABIs, which is resolved through \ul{\bf calling conversion} implemented by collaborative stubs between the guest and host; second, the interleaving of call chains caused by offloaded host functions calling back to the guest, which is tackled via an elegant \ul{\bf emulation reentrancy} at runtime.
Additionally, since switching execution between guest and host modes incurs much higher overhead than direct function calls, we introduce three key \ul{\bf optimization techniques} to mitigate this cost and harvest performance benefits: global reference tables (GRT), fast calling paths (FCP), and partial function outlining (PFO). 
Beyond applications, \NAME can also be applied to shared libraries, enabling speedups for downstream binaries without modification to them. This benefits closed-source applications, and is particularly suitable for commercial software who uses open-source basic libraries, thus holding strong practical value.

To validate the effect of \NAME, we implemented a prototype tool set, \FM, based on the state-of-the-art open-source infrastructures, LLVM\cite{lattnerLLVMCompilationFramework2004} and QEMU\cite{bellardQEMUFastPortable2005}.
\FM consists of the compile-time and runtime tools, supporting the acceleration for both emulating x86-64 on AArch64 and vice versa.
Our evaluaton spans the LLVM Test Suite\cite{LlvmLlvmtestsuite2025}, NAS Parallel Benchmarks\cite{NASParallelBenchmarks2024}, along with multiple dynamicly linked real-world applications.
Experimental results demonstrate that, \FM can accelerate execution by up to 13.03x on AArch64 and 18.91x on x86-64, with an \zzc{a geometric mean}{average} of 3.03x (AArch64) and 3.18x (x86-64), validating \NAME's effectiveness in accelerating cross-ISA DBT. All implementation and experimental codes have been made publicly available, and we continue to refine the the system for deployment in large-scale real-world applications.

In summary, this paper makes the following contributions:
\begin{enumerate}[leftmargin=*]
    \item We propose \NAME, a mechanism that combines the compile-time and runtime to offload guest function calls to the host side for native execution, detailing the guest-to-host forwarding and host-to-guest callbacks.
    \item We further introduce three optimizations, GRT, FCP and PFO, that mitigate the high overhead of guest-host switching and deliver significant performance improvements.
    \item We implement \FM, a prototype for \NAME with all the optimizations, and evaluate it using benchmarks and real-world applications. The results show that \FM can achieve speedups of up to an order of magnitude, validating its practical effectiveness in accelerating cross-ISA emulation for DBT.
\end{enumerate}

\section{Background and Motivation}

\subsection{Background}

\subsubsection{Binary Translation}

Binary translation is a fundamental technique for achieving ISA compatibility, and is widely applied in application migration and in expanding the ecosystem of emerging ISAs. The core principle is to translate binary instructions from a guest ISA  into equivalent instructions of a host ISA for execution.
Because programs often exhibit dynamic behavior (e.g., loading external shared libraries at runtime), translation must be performed dynamically. This leads to Dynamic Binary Translation (DBT), which, despite its flexibility, suffers from substanial performance overhead. This inefficiency arises from from two primary sources.
First, semantic mismatches  between guest and host ISAs  often require multiple host instructions to emulate a single guest instruction, directly degrading execution efficiency.
Second, the translation process itself is computationally expensive, and to ensure real-time translation, DBT typically forgoes deep optimizations during translation, further limiting performance.
Consequently, cross-ISA emulation via DBT can be dozens of times slower than the native execution, severely restricting its applicability in real-world scenarios.

\subsubsection{Cross-compilation}\label{sec:cross-compilation}

Given DBT's inherent limitations, cross-compilation is often considered as a superior alternative for migrating applications to new ISAs. Nowadays, most applications are developed in high-level languages like C and C++, whose source code is largely ISA-agnostic. It is thus theoretically possible to recompile these applications directly into native binaries for the target ISA via the native compilation toolchain.
In practice, however, many applications are not fully target-independent. Their source codes often embed components tightly coupled to the original ISA, mainly manifesting in the following scenarios:
\begin{itemize}[leftmargin=*]
\item Target-specific macros: source code frequently contains ISA-related snippets conditionally compiled via preprocessor macros (e.g., \texttt{\#ifdef x86-64}). These macros are typically used to leverage ISA-specific features such as register layouts, instruction set extensions. If the new ISA is not explicitly supported, cross-compilation may yield missing functionalities or incorrect behaviors.
\item Target-specific assembly code: to pursue performance or perform low-level operations,  applications often include ISA-specific assembly code. Such code can only be executed on the original ISA and is typically rejected by cross-compilation toolchains, preventing a successful build.
\item Unavailable dependencies: even when the application itself is ISA-agnostic,  its direct or indirect dependencies may not be available on the target ISA. This issue is further exacerbated for closed-source commercial libraries that lack support for the new architecture.
\end{itemize}

For these scenarios, existing compiler toolchains often fall into an "all or nothing" dilemma — either extensive source code modifications must be performed to eliminate ISA dependencies, or cross-compilation must be abandoned which falls back to DBT emulation with its prohibitive overhead.

\subsection{Opportunities and Challenges}

\begin{figure}[h]
    \centering
    \includegraphics[width=\linewidth]{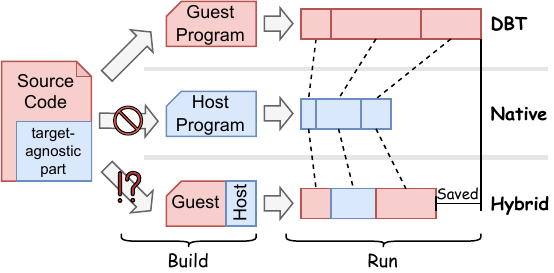}
    \caption{Though the whole source code can't be cross-built to the host target, the target-agnostic part of the source may still be exploited to accelerate DBT.}
    \label{fig:motivation}
\end{figure}

\subsubsection{Coordinated Compilation and Emulation}

The independence inherent in most ISA-specific applications' code bases has not been fully explored. Recent studies \cite{gouicemRisottoDynamicBinary2022, spinkAcceleratingSharedLibrary2024, tanUsingLocalLibrary2018, wangEnablingCrossISAOffloading2017} have studied bypassing DBT by offloading certain guest function calls to the host semantically equivalent functions, thereby significantly accelerating the emulation speed.
For instance, one recent work reported up to a 28x speedup \cite{spinkAcceleratingSharedLibrary2024}. However, existing offloading methods are largely constrained: they either rely on built-in helper functions \cite{tanUsingLocalLibrary2018} or require developers to manually register functions via Interface Definition Language (IDL) \cite{gouicemRisottoDynamicBinary2022, spinkAcceleratingSharedLibrary2024}. Such approaches demand in-depth knowledge of both the ISA and the target functions, greatly limiting practicality require developers to be very familiar with the ISA and the offloaded functions, making it difficult to be applied to other code bases. As a result, their applicability remains confined to a narrow set of standard or common libraries, such as libc, Sqlite, rather than extending to broader, real-world code bases.

From the DBT perspective, the fundamental difficulty in function offloading lies in the loss of key semantic information at the binary level. Compilation and optimization strip away crucial metadata from the source,  such as function signatures, parameter and return value types, all of which are essential for smooth cross-ISA invocation.
For example, without determining the types of functions, it is difficult to correctly pass parameters and convert stack frame layout between the guest ISA and the host ISA, and thus impossible to safely offload function calls from the emulation to native execution.

A promising alternative is to shift offloading to a source-level perspective. At the source code level, full type information and structural semantics are preserved, enabling accurate and reliable identification of offloadable functions.
Through static analysis,  ISA-agnostic code blocks can be cross-compiled to the host ISA, while ISA-specific code is compiled as usual to the original guest ISA.
The final program thus retains full functionality on the guest ISA, but with performance-critical portions offloaded to native host execution. This insight motivates the design of \ul{\bf coordinated compilation and emulation} ({\it Hybrid} in Figure \ref{fig:motivation}) to enable automated and generalizable function offloading, which makes it applicable to diverse code bases without requiring intrusive manual efforts to fully automate such offloading and generally expand its usage to all code bases.

\subsubsection{Challenges of Guest-Host Invocations and Callbacks}

Realizing automated and generalizable source-guided offloading is far from straightforward. It introduces two fundamental challenges that must be addressed to ensure correctness, robustness, and performance:

\textbf{Guest-Host ABI Discrepancies}.
The functions in source code have the same parameter and return value types regardless the ISAs, laying a foundation for cross-ISA interaction. But directly moving guest functions to the host side is not feasible.
The core obstacle lies in the Application Binary Interfaces (ABIs) of different ISAs. The ABI governs low-level function call conventions, including parameter-passing, register usage, and stack frame layout \cite{AbiaaAapcs64Main2025, X86PsABIsX86642025}.
Without proper calling conversion, offloaded execution risks runtime failures such as incorrect parameter delivery, stack corruption, or misaligned return values. Existing approaches \cite{gouicemRisottoDynamicBinary2022, spinkAcceleratingSharedLibrary2024, tanUsingLocalLibrary2018} often rely on manual intervention to handle such conversions, which is error-prone and impractical at scale. In addition, functions usually reference external globals (e.g., variables, constants, or other functions). These globals must be properly propagated to the host side to preserve semantic correctness during offloaded execution, adding another layer of complexity to the calling conversion process.

\textbf{Host-to-Guest Callbacks and Emulation Reentrancy}.
Offloaded host functions are not always self-contained; they may need to call back guest functions like invoking function pointers passed from the guest.
This necessitates establishing a bidirectional calling channel between host-native code and guest-emulated code. Current solutions typically restrict themselves to callback-free scenarios, such as leaf functions that do not invoke others. For complex call chains involving callbacks remain unsolved in general. Handling these scenarios requires carefully managing context switches between emulated and native execution modes while preserving runtime state consistency. Beyond callbacks, non-local control flows in general-purpose code (e.g. \texttt{longjmp}, exceptions) introduce additional complications. These constructs disrupt standard call-return semantics, making cross-ISA invocation particularly error-prone and difficult to generalize.

A combination of code compilation and runtime support is critical to tackle the aforementioned challenges. Moreover, since crossing guest-host boundaries inevitably incurs overhead, optimizations are necessitate to mitigate performance penalties and share key insights into its implementation.

\section{Design}

\subsection{Overview}

\begin{figure}[b]
    \centering
    \includegraphics[width=\linewidth]{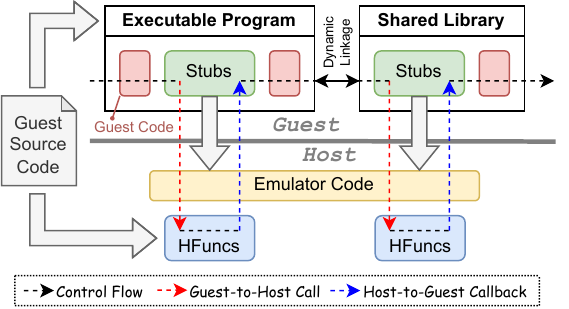}
    \caption{The overview of \NAME mechanism.}
    \label{fig:overview}
\end{figure}

Figure \ref{fig:overview} outlines the general framework of a proposed mechanism. To migrate specific function executions from an emulated environment to a native host environment, coordinated support is required during both compilation and runtime phases. 

In the compilation stage, suitable target functions are identified, with their core logic extracted and adapted for native host execution. Since the host's ABI differs from the emulated environment, intermediate bridging stub functions are retained in the emulated side to manage cross-environment function call conversions. Extracted functions may include references to other components in the emulated environment, requiring reverse conversion support for callback scenarios.

At runtime, the stub functions handle call conversion and triggers environment switching via dedicated interface calls, which transfer execution control to the host-side adapted functions. Interleaved call chains from host to emulated executions may arise from cross-environment callbacks, necessitating support for nested environment switching. A complementary interface is introduced to facilitate switching back to the host environment upon callback completion. Critical requirements include non-intrusiveness, ensuring the mechanism does not disrupt normal emulation operations.

This mechanism is also applicable to library-level functions, enabling performance optimization by migrating library function calls. For dynamically linked libraries, optimization can be achieved by replacing the library binaries directly, without modifying or recompiling the target application. Binary compatibility of shared libraries ensures functional correctness after updates, allowing this approach to benefit both open-source and closed-source applications. This feature is particularly useful for scenarios involving commercial software integrated with standard libraries, offering practical applicability.

\subsection{Calling Conversion}

Calling conversion is jointly handled by stub functions in the emulated environment and offloaded functions derived from original components, instead of relying solely on host-side binary adjustments. The latter would bind adapted functions tightly to specific emulated architectures and internal emulator designs. By allocating part of the conversion task to the emulated side, the proposed mechanism achieves a concise, versatile, and maintainable interface.

In the emulated environment, the original function implementation is replaced with bridging code. When the function is invoked, the stub prepares arguments and other things and send them to the offloaded functions on the host side. This data exchange method allows adapted functions to operate independently of guest or host ISAs, ensuring architecture neutrality and implementation simplicity. A similar approach applies to data exchange for cross-environment callbacks, though a key distinction exists: guest functions called indirectly (e.g., via function pointers) lack built-in unpacking capabilities, requiring additional bridging code in the emulated environment. Adapted functions may reference global elements (including required callback components) that reside in the emulated address space and must be transmitted to the host.

\subsection{Emulation Reentrancy}

Host-executed functions may call-back emulated components, creating an "emulation reentrancy" issue. This interacts with QEMU’s runtime stack layouts. We extend QEMU’s built-in switching mechanism to manage cross-environment transitions. To fix this, we build host function contexts on the guest stack and the host stack frames, keeping stack consistency and matching the emulated program’s logic. This mechanism supports both guest-to-host calls and host-to-guest callbacks.

Auxiliary functions isolate host and emulated execution, ensuring strong compatibility. Even if the emulated program crashes, QEMU stays stable—host functions have no unintended side effects on QEMU’s native operations. LLVM IR resolves underlying interface differences, making it the ideal layer for implementing this mechanism—compiled host code reliably operates on emulated data.

\subsection{Optimizations}

Cross-environment calls are costly—only long functions benefit. To optimize for short functions, we introduce three key methods.

\textbf{Global Reference Table (GRT):} Basic design incurs unnecessary construction of same data for each cross-side function calls. GRT pre-stores them in global constants to eliminates those costs.

\textbf{Fast Calling Path (FCP):} FCP lets offloaded functions call each others directly without switching to the guest emulation. It can effectively reduce cross-boundary overhead.

\textbf{Partial Function Outlining (PFO):} PFO expands offloadable functions, making originally un-offloadable functions offloadable. Hence we support offloading code using variadic calls and other cases. For context-sensitive code, its complement is split instead of themselves. 

\section{Evaluation}

\subsection{Experimental Methodology}

\subsubsection{Testbed}
We evaluate \FM, the \NAME prototype, on both x86-64 and AArch64 machines. The detailed configurations are as listed in Table \ref{tab:testbed}.

\begin{table}[h]
    \caption{Configurations of testbed.}
    \centering
    \begin{tabular}{l|c|c}
    \toprule
            & \textbf{x86-64}
            & \textbf{AArch64}
            \\
    \hline
        \textbf{CPU}
            & \makecell{AMD Ryzen 9 \\ 5950X @ 3.4GHz}
            & \makecell{Phytium \\ FT-2000+/64 @ 2.2GHz}
            \\
    \hline
        \textbf{RAM}
            & 32GB
            & 128GB
            \\
    \hline
        \textbf{OS}
            & \multicolumn{2}{c}{Ubuntu 24.04}
            \\
    \hline
        \textbf{LLVM/Clang}
            & \multicolumn{2}{c}{18.1}
            \\
    \hline
        \textbf{QEMU}
            & \multicolumn{2}{c}{9.1.50 (modified)}
            \\
    \bottomrule
    \end{tabular}
    \label{tab:testbed}
\end{table}

\subsubsection{Workloads}

Our evaluation assesses monolithic benchmark programs, where \FM is applied directly to their source codes for offloading functions. Table \ref{tab:cases-1} summarizes the full list of workloads, which include three customized micro-benchmarks, four randomly chosen C programs from the LLVM test suite \cite{LlvmLlvmtestsuite2025}, and the complete set of NAS parallel benchmarks \cite{NASParallelBenchmarks2024}.

\begin{table}[h]
    \caption{Evaluated workloads (Wkld).}
    \centering
    \begin{tabular}{c|c|}
    \toprule
    \textbf{Source} & \textbf{Wkld} \\
    \midrule
    \multirow{3}{*}{\makecell{Customized}}
        & matpowsum \\
        & cjson \\
        & lua \\
    \hline
    \multirow{4}{*}{\makecell{LLVM\\test-suite}}
        & obsequi \\
        & oggenc \\
        & sgefa \\
        & viterbi \\
    \bottomrule
    \\
    \end{tabular}
    \begin{tabular}{|c|c|c}
    \toprule
    \textbf{Source} & \textbf{Name} & \textbf{Wkld} \\
    \midrule
    \multirow{8}{*}{\makecell{NAS\\Parallel\\Bench}} 
        & BT & npbbt \\
        & CG & npbcg \\
        & EP & npbep \\
        & FT & npbft \\
        & LU & npblu \\
        & MG & npbmg \\
        & SP & npbsp \\
        & IS & npbis \\
    \bottomrule
    \end{tabular}
    \label{tab:cases-1}
\end{table}

\newcommand{\Snative}{\texttt{native}\ }
\newcommand{\Sqemu}{\texttt{qemu}\ }
\newcommand{\Sbasic}{\texttt{TECH}\ }
\newcommand{\Sgrt}{\texttt{TECH-g}\ }
\newcommand{\Sfcp}{\texttt{TECH-gf}\ }
\newcommand{\Spfo}{\texttt{TECH-gfp}\ }
\subsubsection{Metrics}
We evaluate our system using two primary metrics: 1) performance, and 2) program size. The following schemes are compared:
\begin{itemize}[leftmargin=*]
    \item \Snative: programs built and executed natively.
    \item \Sqemu: original guest programs emulated with the vanilla QEMU on the host machine.
    \item \Sbasic: \FM with baseline implementation of \NAME, i.e., without any enabled optimizations.
    \item \Sgrt: \FM using \NAME enabled with Global Reference Table (GRT) optimization.
    \item \Sfcp: \FM using \NAME with both GRT and Fast Calling Path (FCP) optimizations.
    \item \Spfo: \FM using \NAME with all optimizations: GRT, FCP and Partial Function Outlining (PFO).
\end{itemize}

\subsection{Performance Improvement}\label{sec:performance}

\begin{figure*}[t]
    \centering
    \centering
    \includegraphics[width=\linewidth]{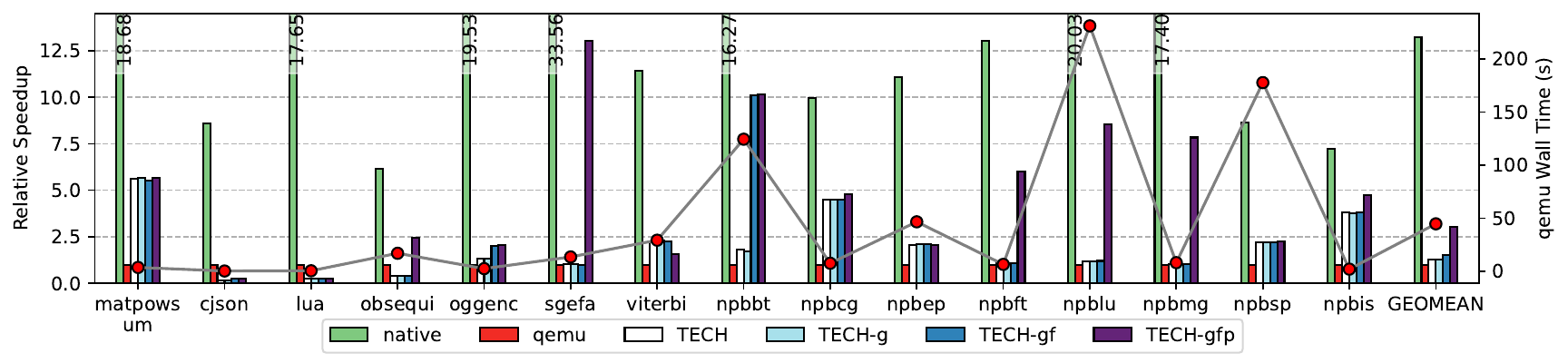}
    \caption{Performance of the x86-64 emulation on AArch64. The bars (left y-axis) show speedup relative to original QEMU, with the folding line (right y-axis) showing its wall time.}
    \label{fig:exp1}
\end{figure*}

\begin{figure*}[t]
    \centering
    \includegraphics[width=\linewidth]{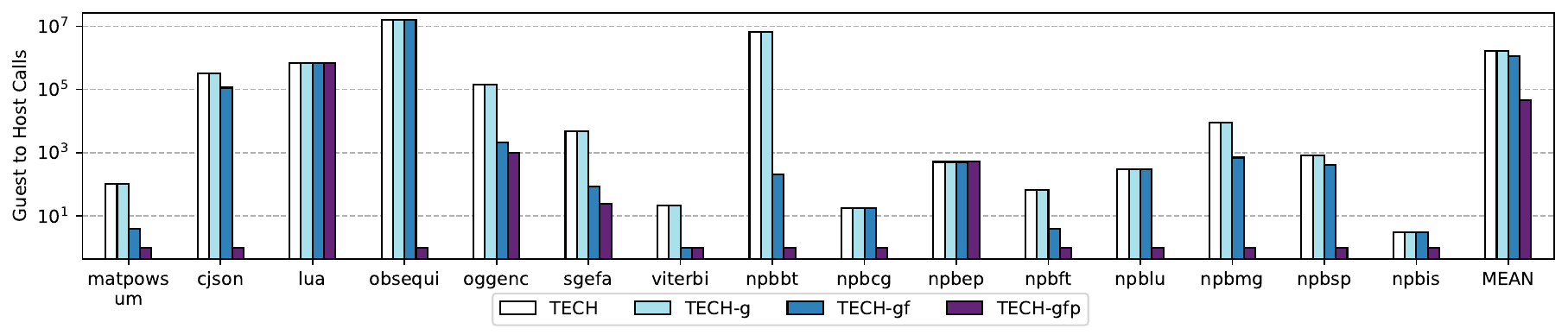}
    \caption{Number of guest-to-host calls during the execution of all the workloads.}
    \label{fig:callnum}
\end{figure*}

Figure \ref{fig:exp1} presents the measured speedup (left y-axis) and wall time (right y-axis) for all workloads on the AArch64 machine. It can be seen that the \Sqemu emulation is significantly slower than \Snative execution with a geometric mean slowdown of 13.23x.
When applying \FM along with all its optimizations (i.e., \Spfo), the emulation performance is improved by a geometric mean of 3.03x (up to 13.03x).

A closer examination of the data in the figure reveals several interesting observations. First, the baseline version of \FM (\Sbasic) achieves a geometric mean speedup of 1.29x, while the sole GRT optimization (\Sgrt) does not lead to a significant change (still around 1.29x). This indicates that the primary overhead of cross-guest-host invocations does not lie in the arguments and return value conversions, but rather in the internal works of QEMU, including system call handling, context switching, and others. Consequently, reducing the number of guest-host boundary crossings (\Sfcp) can significantly mitigate overhead, which is corroborated by the optimization effect of FCP (1.54x).

Second, it can be observed that most significant speedup comes from the final PFO optimization. This indicates that there were indeed many hot functions containing problematic instructions, which prevent them from being offloaded to the host side. Our in-depth investigation of these applications reveals that calls to variadic functions are the primary cause. Source codes often include safety check statements using functions like \texttt{printf}, which hinder the offloading of the entire function but are usually not triggered at runtime. The PFO optimization effectively addresses this case.

Third, negative optimizations do exist: in the two workloads \texttt{cjson} and \texttt{lua}, \FM results in worse performance even with all optimizations being enabled. This is because the two applications contain a large number of short functions that call each other. Offloading them not only fails to bring much speedup but also causes frequent crossings of the host-guest boundary. This insight conveys that the offloading is not a guaranteed gain; we must wisely select which function to offload. Our current prototype adopts a very simple strategy  of filtering out functions whose number of basic blocks and instructions exceeding a certain threshold. More sophisticated strategies are possible, such as better cost models and profiling, which we leave for future work. However, it must be noted that \NAME offloading is incremental and allows degradation to pure emulation in the worst-case scenario. Therefore, the practical performance of \FM can definitely be no worse than that of the original \Sqemu.

\subsection{Function Statistics}

\subsubsection{\NAME Invocation Count}

Figure \ref{fig:callnum} gives the count of guest-to-host calls during the execution of all workloads in Figure \ref{fig:exp1}. It's quite obvious that there is a strong positive correlation between these two sets of statistics. This insight confirms the conjecture that the overhead of \NAME mainly stems from crossing the host-guest boundary.

Just as expected, the GRT optimization poses no effect to the invocation count, whereas the FCP optimization significantly reduces the invocation number. For instance, the count drops from 6,713,003 to 206 in the \textit{npbbt} workload. However, greater effect comes from its conjunction with the PFO optimization, where the number decreases by more than an order of magnitude. Especially, 11 workloads triggers only one single \NAME invocation after PFO optimization. \zzc{}{This implies that almost the entire program is executed at the host side. These results confirm the effectiveness and necessity of the optimizations.}

One abnormal point is \textit{cjson}. After PFO optimization, its invocation count drops to one, but the performance remains to be inferior to vanilla \Sqemu, though it does show improvement compared to the baseline \Sbasic (1.85x speedup). After investigation, we attribute this to the callbacks to \textit{libc} functions which is not offloaded by \FM. Since most \textit{cjson} functions are very light with little computation, the time saved by native execution is far less than that spent on callbacks. This finding inspires us to preferably apply \NAME to the lower half of the entire software stack in a bottom-up manner to eliminate host-to-guest callbacks as much as possible.

\subsubsection{Function Offloading Coverage}

\begin{figure}[b]
    \centering
    \includegraphics[width=0.95\linewidth]{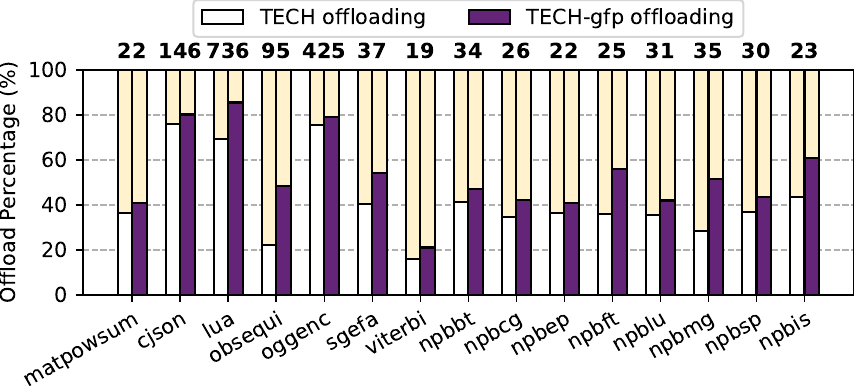}
    \caption{Function offloading coverage results. The values above the bars are the total function numbers.}
    \label{fig:funcnum}
\end{figure}

The function coverage results show how many functions are offloaded by \FM. The more function offloaded, the higher chance that host-to-guest callbacks may be eliminated with by the FCP optimization. Figure \ref{fig:funcnum} clearly shows that PFO increases the function coverage. For instance, the number of offloaded functions increased from 21 to 46 in the \textit{obsequi} workload, which reduces \NAME invocations from 16,206,473 to just 1, yielding a 6.34× speedup compared to the baseline \Sbasic.

There is no strong correlation between function coverage and performance improvement. This is because the additionally offloaded functions may not be hot or even used. A typical example is the \textit{lua} workload, where a significant improvement in function coverage doesn't bring any changes to its performance results. This inspires us to explore the combination of profiling methods to selectively offload hot functions in the future.

\subsection{Additional Studies}

\subsubsection{Emulate x86-64 on AArch64}

\begin{figure}[b]
    \centering
    \includegraphics[width=\linewidth]{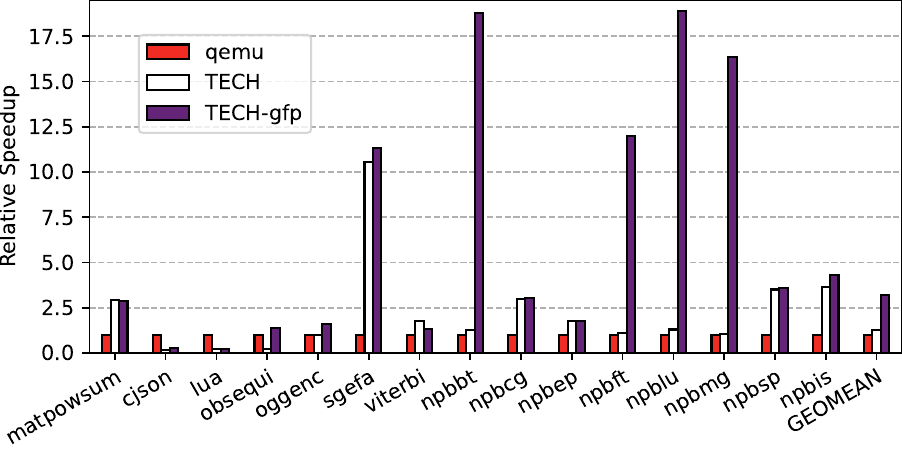}
    \caption{Relative performance of the AArch64 emulation on x86-64.}
    \label{fig:x86}
\end{figure}

The universal design of \NAME makes it easy to be extended to other guest/host ISA combinations. So, without much efforts, we additionally echo the performance of emulating AArch64 on the x86-64 machines, with the results shown in Figure \ref{fig:x86}. The performance trend on x86-64 is consistent with that on AArch64 with a slightly better geometric mean(3.18x) and maximum (18.91x). This result fully demonstrates the stability and reliability of \NAME, and also reflects its low sensitivity to specific ISA combinations.

\subsubsection{Effects on Shared Library}

As mentioned before, \NAME is not limited to applications but can also be extended to the libraries. To this end, we supplement acceleration experiments targeting two open-source libraries: \textit{libpng} and \texttt{zlib}, along with four applications — \textit{apng2gif}, \textit{optipng}, \textit{imagemagick}, and \textit{zlib-flate} — to test the actual effect of offloading library functions. It is worth noting that all applications are pre-built binaries directly downloaded from the Ubuntu APT repository and dynamically linked. So the acceleration is performed solely by replacing the corresponding library files, without modification to the applications themselves. The relevant experimental results are shown in Table \ref{tab:so-apps}.

It is evident that accelerating the \textit{zlib} library significantly performs better than that of \textit{libpng}. Moreover, the acceleration effects of different libraries are additive: in the scenario where \textit{imagemagick} converts compressed PNG images, accelerating both \textit{libpng} and \textit{zlib} altogether achieves an overall performance improvement of 3.96x, which is higher than the effects of accelerating either library alone (3.87x and 1.20x). These results indicate that \NAME can still improve the overall performance of applications by optimizing the dynamic libraries they depend on, even when the application source code is unavailable. What's more, library-level acceleration is universal, i.e., once a basic library is optimized, all applications relying on it can benefit. This significantly expands the scope of application and influence of \NAME. 

\section{Conclusion}

Conventional DBT incurs substantial emulation overhead, while existing cross-compilation techniques often adhere to an "all-or-nothing" paradigm, rendering them impractical for legacy or closed-source software with target-specific elements.
To address the challenges, we propose \NAME, a function offloading mechanism that coordinates compile-time and runtime supports. \NAME extracts eligible guest functions, generates their equivalent host versions, and forwards calls to them, thus bypassing expensive DBT emulation.
The design resolves ABI discrepancies via calling conversion and ensures correct execution across interleaved chains through emulation reentrancy. Furthermore, our optimizations (GRT, FCP, PFO) effectively reduce guest-host switching costs.
We implement the ideas in \FM, a prototype based on LLVM and QEMU. Evaluation results demonstrate substantial performance improvements, achieving up to 13.03x speedup on AArch64 and 18.91x on x86-64, with geometric means of 3.03x and 3.18x, respectively. These results validate the effectiveness and robustness of \NAME in accelerating DBT.
We are working on applying \NAME to real-world application workloads in large scale currently. Looking forward, we plan to explore deeper compiler-emulator co-optimizations, and more adaptive offloading strategies guided by workload characteristics.

\begin{table}[t]
    \centering
    \caption{Shared library acceleration}
    \begin{tabular}{c|c|c}
    \toprule
    {\bf Offloaded Lib.} & {\bf Downstream App.} & {\bf Speedup} \\
    \hline
    \multirow{3}{*}{libpng}
        & apng2gif & 1.04x \\
        & optipng & 1.05x \\
        & imagemagick & 1.20x \\
     \hline
     \multirow{2}{*}{zlib}
        & imagemagick & 3.87x \\
        & zlib-flate & 16.48x \\
    \hline
    libpng+zlib
        & imagemagick & 3.96x \\
    \bottomrule
    \end{tabular}
    \label{tab:so-apps}
\end{table}

\balance
\newpage

\bibliographystyle{ACM-Reference-Format}
\bibliography{ref}

\appendix

\end{document}